\documentclass[aps,prl,floatfix,12pt,superscriptaddress]{revtex4-1}

\makeatletter
\newcommand\thefontsize[1]{{#1 The current font size is: \f@size pt\par}}
\makeatother

\usepackage{siunitx}
\usepackage{verbatim}
\usepackage{graphicx}
\usepackage{nicefrac}

\begin{document}

\title{Arrays of individually controllable optical tweezers based on 3D-printed microlens arrays}

\author{Dominik Sch\"{a}ffner} 
\author{Tilman Preuschoff}
\affiliation{Institut f\"{u}r Angewandte Physik, Technische Universit\"{a}t Darmstadt, Schlossgartenstra\ss e 7, 64289 Darmstadt, Germany}
\author{Simon Ristok}
\affiliation{4.~Physikalisches Institut, Universit\"{a}t Stuttgart, Pfaffenwaldring 57, 70569 Stuttgart, Germany}
\author{Lukas Brozio}
\author{Malte Schlosser}
\affiliation{Institut f\"{u}r Angewandte Physik, Technische Universit\"{a}t Darmstadt, Schlossgartenstra\ss e 7, 64289 Darmstadt, Germany}
\author{Harald Giessen}
\affiliation{4.~Physikalisches Institut, Universit\"{a}t Stuttgart, Pfaffenwaldring 57, 70569 Stuttgart, Germany}
\author{Gerhard Birkl}
\affiliation{Institut f\"{u}r Angewandte Physik, Technische Universit\"{a}t Darmstadt, Schlossgartenstra\ss e 7, 64289 Darmstadt, Germany}

\email{\authormark{*}dominik.schaeffner@physik.tu-darmstadt.de} 

\homepage{www.iap.tu-darmstadt.de/apq} 
\date{\today}

\begin{abstract}
We present a novel platform of optical tweezers which combines rapid prototyping of user-definable microlens arrays with spatial light modulation (SLM) for dynamical control of each associated tweezer spot. Applying femtosecond direct laser writing, we manufacture a microlens array of 97 lenslets exhibiting quadratic and hexagonal packing and a transition region between the two. We use a digital micromirror device (DMD) to adapt the light field illuminating the individual lenslets and present a detailed characterization of the full optical system. In an unprecedented fashion, this novel platform combines the stability given by prefabricated solid optical elements, fast reengineering by rapid optical prototyping, DMD-based real-time control of each focal spot, and extensive scalability of the tweezer pattern. The accessible tweezer properties are adaptable within a wide range of parameters in a straightforward way.\\
 \\
\bfseries{Original Citation (Open Access):\\
Optics Express 28, 8640-8645 (2020); DOI: 10.1364/OE.386243}
\end{abstract} 
\maketitle

\section{Introduction}
With his seminal publications on optical tweezers in the 1970s \cite{Ashkin1970,Ashkin2006}, A. Ashkin has put forth a technique which nowadays is widely used in many different research fields, including physics, chemistry, biology, and medical research \cite{Ashkin1980,Ashkin2006,Polimeno2018,Dhakal2018}. Granting a high level of dynamic range in the interaction strength between light and matter, optical tweezers are extremely versatile and can be used in order to monitor and control the external degrees of freedom of a large variety of target objects ranging from single atoms to living organisms. Surpassing the original configuration of using one single laser spot, means of acousto-optical deflection (AOD) \cite{Akselrod2006,Endres2016}, spatial light modulation (SLM) \cite{Akselrod2006,Curtis2002,Bergamini2004,Ferrari2005,Lee2006,Nogrette2014,Stuart2018}, or implementations based on microlens arrays (MLA) \cite{Schlosser2011,OhldeMello2019} facilitate multi-site tweezer systems with complex and parallelized means of manipulation. The MLA approach, in specific, offers unprecedented scalability due to the lithographic production process. This allows for the creation of many thousands of diffraction-limited tweezer sites \cite{Schlosser2011,OhldeMello2019}, not limited by the finite frequency spectrum of AODs or the resolution-versus-size constraints of liquid-crystal-based SLMs.  However, a modification of the tweezer pattern requires a new fabrication of the employed MLA with large investments for the lithographic masks. This causes long turnaround times and reduces flexibility.

In this work, we report on the realization of a novel platform of optical tweezers that overcomes previous limitations by combining two powerful state-of-the-art technologies: rapid prototyping of microlens arrays based on direct laser writing and in-situ access and dynamic control of each tweezer spot based on digital micromirror device (DMD) spatial light modulation. 
We achieve an efficiency between the values reported for holographic DMD tweezer generation \cite{Holland2017} and liquid-crystal (LC) based phase modulating  SLMs \cite{Nogrette2014}, but our DMD-based approach offers at least an order of magnitude higher addressing rates as compared to LC-based configurations \cite{Nogrette2014,Holland2017}. In this way, we significantly extend the state-of-the-art of optical tweezers by combining the high-quality foci typical for microlens setups with kilohertz rates for addressability in versatile yet stable configurations. We present in detail design and manufacturing of the MLA, the optical setup combining the MLA with the DMD-based control of each individual tweezer, and the characterization of the full optical setup. This novel experimental platform finds immediate application in the investigation and manipulation of biological systems, dielectric particles, large atomic ensembles, as well as individual neutral atoms. For a detailed discussion, we have adapted the system to single-atom quantum simulation as demonstrated in \cite{Schlosser2011,OhldeMello2019,Sturm2017} with design parameters optimized for a setup similar to the one detailed in \cite{OhldeMello2019}.

\begin{figure}[b]
	\centering\includegraphics[width=0.77\textwidth]{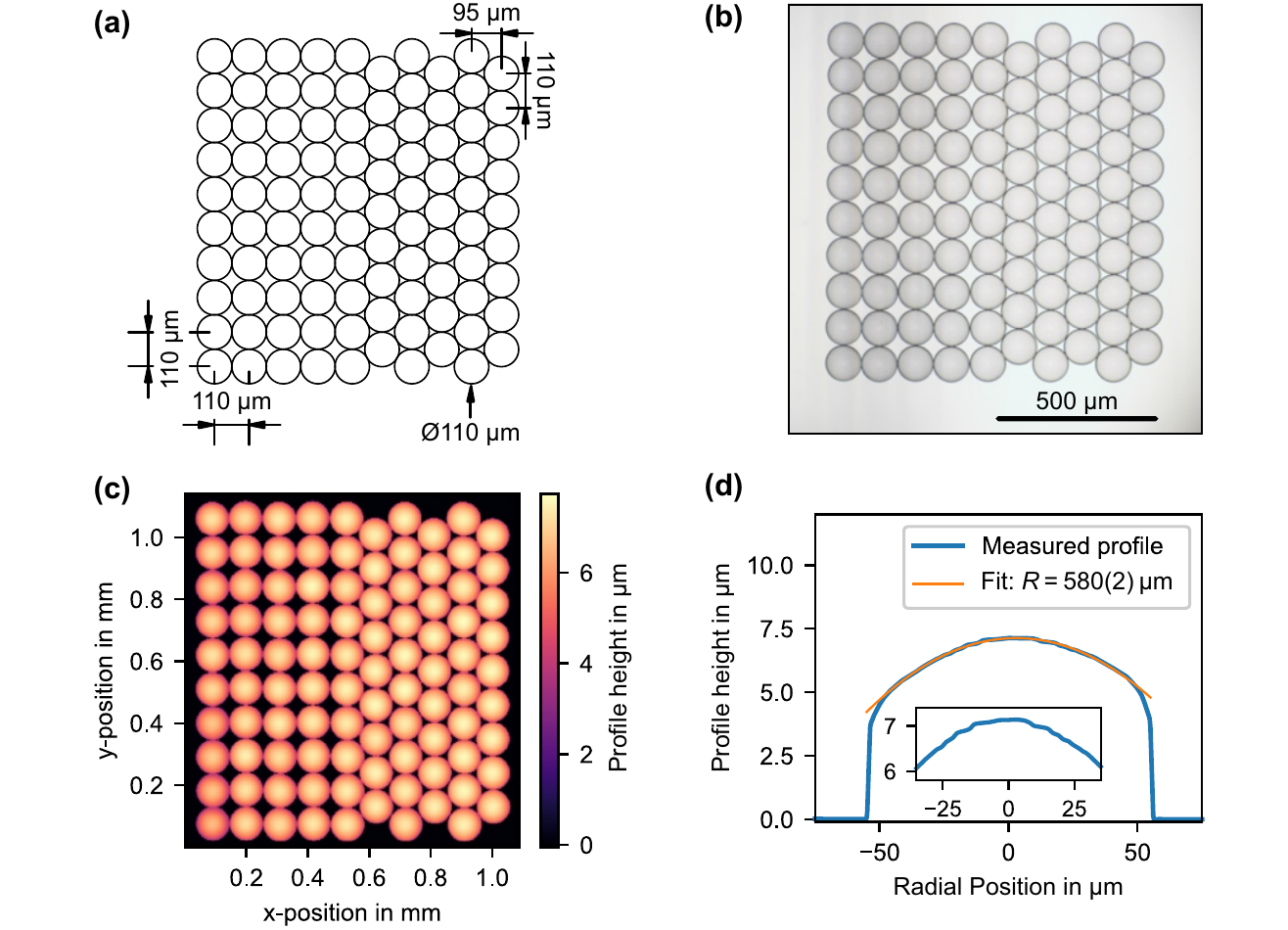}
	\caption{
	Layout and characterization of the manufactured microlens array.
	(a) The array is designed to consist of 97 lenslets with a fundamental pitch of \SI{110}{\micro\meter} in quadratic and hexagonal packing including a region of transition from rectangular to hexagonal symmetry.
	(b) Inspection of the manufactured array by white-light microscopy confirms the successful manufacturing process.
	(c) A confocal microscope is used to analyze the surface of the microlens array.
	(d) Profile of one of the microlenses and fit for the determination of the quality of the spherical surface and the radius of curvature $R$. The inset provides a more detailed view on the central region of the lenslet.\label{MLA_design}}
\end{figure}
\section{Design and manufacturing of microlens arrays with versatile geometries}
\label{sec_manufacture}
Three-dimensional printing by femtosecond direct laser writing of microfabricated optics has become a well established technology in recent years \cite{Kawata2001,Gissibl2016,Yuan2019} providing tremendous accuracy and short design and production cycles. We apply this technique to the production of a custom-design MLA consisting of circular shaped spherical lenslets. Figure \ref{MLA_design} presents the layout (a) of the microlens array together with the results of the post-fabrication characterization measurements (b-d). Maintaining a lens pitch and diameter of $\SI{110}{\micro\meter}$, the array features sections with quadratic and hexagonal packing, respectively, including a region of transition from rectangular to hexagonal symmetry. We use 3D dip-in direct laser writing \cite{Kawata2001,Gissibl2016} with a femtosecond lithography system (Nanoscribe Photonic Professional GT) to sequentially imprint 97 spherical microlenses on a fused silica substrate with a thickness of \SI{700}{\micro\meter}. As printing can be executed with a writing speed of about 10 lenslets per hour this manufacturing approach is especially suited for rapid prototyping of systems with up to about \SI{100}{} lenslets. If required, processing speed can be significantly increased by parallelized writing techniques \cite{Yuan2019}.

A typical height of the curved lens profile measures only a few microns. Hence keeping the surface roughness at a minimum plays an important role. We use the photoresist Nanoscribe IP-S (index of refraction $n=\SI{1.505}{}$ at a wavelength of $\SI{796}{\nano\meter}$ \cite{Schmid2018}) which has been specifically developed for this purpose. The printing resolution of the lithography system normal to the substrate surface is $\SI{100}{\nano\meter}$.
As a first step of verifying a successful fabrication process, we inspect the printed lens array with a standard optical microscope (Nikon Eclipse LV100) and retrieve the white light image shown in Fig.~\ref{MLA_design}(b).
Using a commercial confocal microscope setup (NanoFocus $\mathrm{\mu}$surf expert), we obtain a detailed view of the microlens surfaces (Fig.~\ref{MLA_design}(c)). From a typical lens profile (Fig.~\ref{MLA_design}(d)) we extract a radius of curvature of $R=\SI{580(2)}{\micro\meter}$ of the spherical surface with an rms-deviation of \SI{0.03}{\micro\meter} from a perfect sphere within \SI{90}{\percent} of the lens diameter (i.e. \SI{99}{\micro\meter}). This results in an effective numerical aperture of $\mathrm{NA}=\SI{0.045(2)}{}$.
The inset in Fig.~\ref{MLA_design}(d) provides an enlarged view of the profile. The surface irregularities match the \SI{100}{\nano\meter} resolution of the lithography system normal to the substrate surface. As a next step, we analyze the focal plane of the microlens array when illuminated by a laser beam at a wavelength of $\lambda=\SI{796.7}{\nano\meter}$ and a diameter much larger than a microlens diameter. Approximating the resulting foci with two-dimensional Gaussian intensity distributions yields a $\nicefrac{1}{e^2}$-waist of $w_\mathrm{q}=\SI{6.5(3)}{\micro\meter}$ in the quadratic and $w_\mathrm{h}=\SI{6.4(3)}{\micro\meter}$ in the hexagonal region, respectively. These values are in excellent agreement with simulated wavefront propagation using the lenslet design parameters and the measured radius of curvature leading to diffraction limited performance of the tweezer array as detailed in Sec.~\ref{sec_light_field_control}.
\section{Tweezer array with extended light field control using a spatial light modulator}
\label{sec_light_field_control}
\begin{figure}[t]
	\centering
	\includegraphics[width=0.84\textwidth]{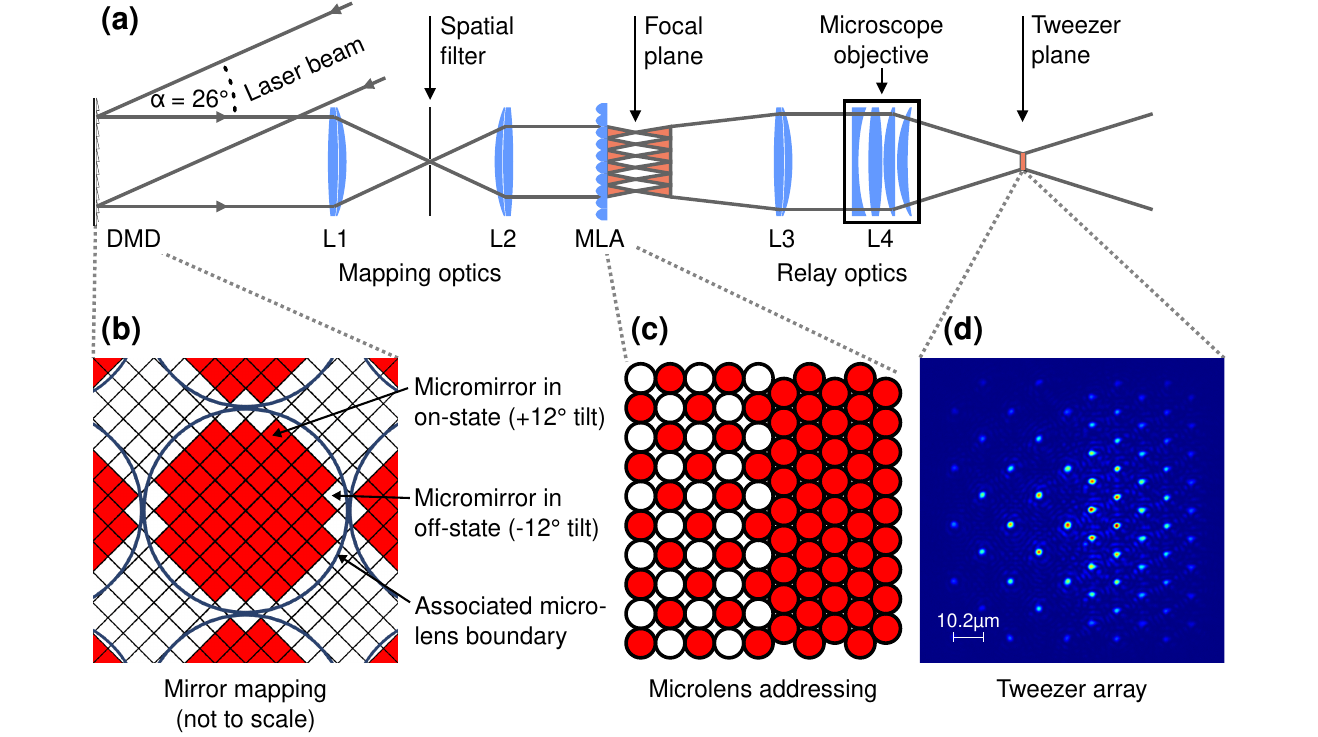}
	\caption{
		Novel tweezer platform including DMD-based microlens addressing, mapping optics, and relay optics for demagnification.
		(a) The incident laser beam illuminates the surface of the DMD under an incident angle of $\alpha\approx\SI{26}{\degree}$ resulting in the third diffraction order to be deflected orthogonally off the DMD surface. Two lenses L1 and L2 map the DMD surface onto the microlens array whose focal plane is demagnified onto the tweezer plane by the relay optics with lens L3 and microscope objective L4. As shown schematically in (b) switching mirrors within circular regions on the DMD can be used to (c) address individual and groups of microlenses and (d) switch the corresponding tweezer foci.\label{fig_optical_setup}}
\end{figure}

The schematic setup of our novel tweezer platform is visualized in Fig.~\ref{fig_optical_setup}(a). We employ a DMD (Texas Instruments Lightcrafter EVM) in order to render the microlenses selectively addressable and to dynamically modify the light field that illuminates every single lenslet of the MLA during operation of the tweezer array. The micromirror array with a size of \SI{6.57}{\milli\meter} x \SI{3.70}{\milli\meter} features 608 x 684 quadratic mirrors with a pitch of \SI{7.64}{\micro\meter} along the mirror edges whose tilt angle relative to the DMD surface can be individually set to two distinct orientations of $\pm\SI{12}{\degree}$, respectively, with an update rate of up to \SI{4}{\kilo\hertz}. For our application \cite{OhldeMello2019}, a Gaussian laser beam at a wavelength of $\lambda=\SI{796.7}{\nano\meter}$ and a $\nicefrac{1}{e^2}$-radius of $w_\mathrm{B}=\SI{1.2}{\milli\meter}$ illuminates the DMD under an incident angle of $\alpha\approx\SI{26}{\degree}$ leading to deflection of the third diffraction order normal to the DMD surface and along the optical axis of the subsequent optical setup. A spatial filter blocks all unwanted diffraction orders. We obtain a typical efficiency of the DMD section of our setup of $\eta=\SI{33}{\percent}$ which is limited by the fact that $\alpha$ deviates from the angle of specular reflection off a single mirror of \SI[retain-explicit-plus]{+24}{\degree}, by light being diffracted into other diffraction orders, by finite mirror and window efficiencies, and a limited filling factor as detailed in the datasheet.

Using a confocal telescope of two achromatic doublets L1 and L2 with focal lengths of  $f_1=\SI{100}{\milli\meter}$ and $f_2=\SI{45}{\milli\meter}$, respectively, the DMD plane is mapped onto the microlens array with every lenslet illuminated by a circular pattern of mirrors as schematically illustrated in Fig.~\ref{fig_optical_setup}(b). Due to the fact that the pitch of the microlens array (\SI{110}{\micro\meter}) is not a multiple of the demagnified micromirror pitch ($\SI{7.64}{\micro\meter} \cdot f_2/f_1 = \SI{3.44}{\micro\meter}$) the numbers of mirrors that are assigned to different lenslets are not identical, yet typically \SI{730(8)}{}. This results in an effective microlens illumination diameter of \SI{109(1)}{\micro\meter}.
As exemplified in Fig.~\ref{fig_optical_setup}(c), irradiation of each microlens, and consequently each focal spot, can be individually turned on (off) by tilting the respective mirrors into the \SI[retain-explicit-plus]{+12}{\degree} (\SI{-12}{\degree}) orientation, respectively. Sections of the MLA not corresponding to a lenslet are not illuminated. Thus, stray light resulting from areas outside the lenslets is avoided. In regions with quadratic (hexagonal) packing, \SI{71}{\percent} (\SI{82}{\percent}) of the DMD mirrors contribute to the tweezer light field.

The focal plane of the microlens array is demagnified (magnification $M_{34}=f_4/f_3=0.094(3)$) using an achromatic doublet L3 with a focal length of $f_3=\SI{400}{\milli\meter}$ in combination with a microscope objective L4 with an effective focal length of $f_4=\SI{37.5(10)}{\milli\meter}$  yielding a tweezer array with a measured spot separation of $d=\SI{10.2(3)}{\micro\meter}$. Figure \ref{fig_optical_setup}(d) shows an image of the tweezer plane taken with a CCD camera after applying the illumination pattern depicted in Fig.~\ref{fig_optical_setup}(c).
\begin{figure}[t]
	\centering\includegraphics[width=0.77\textwidth]{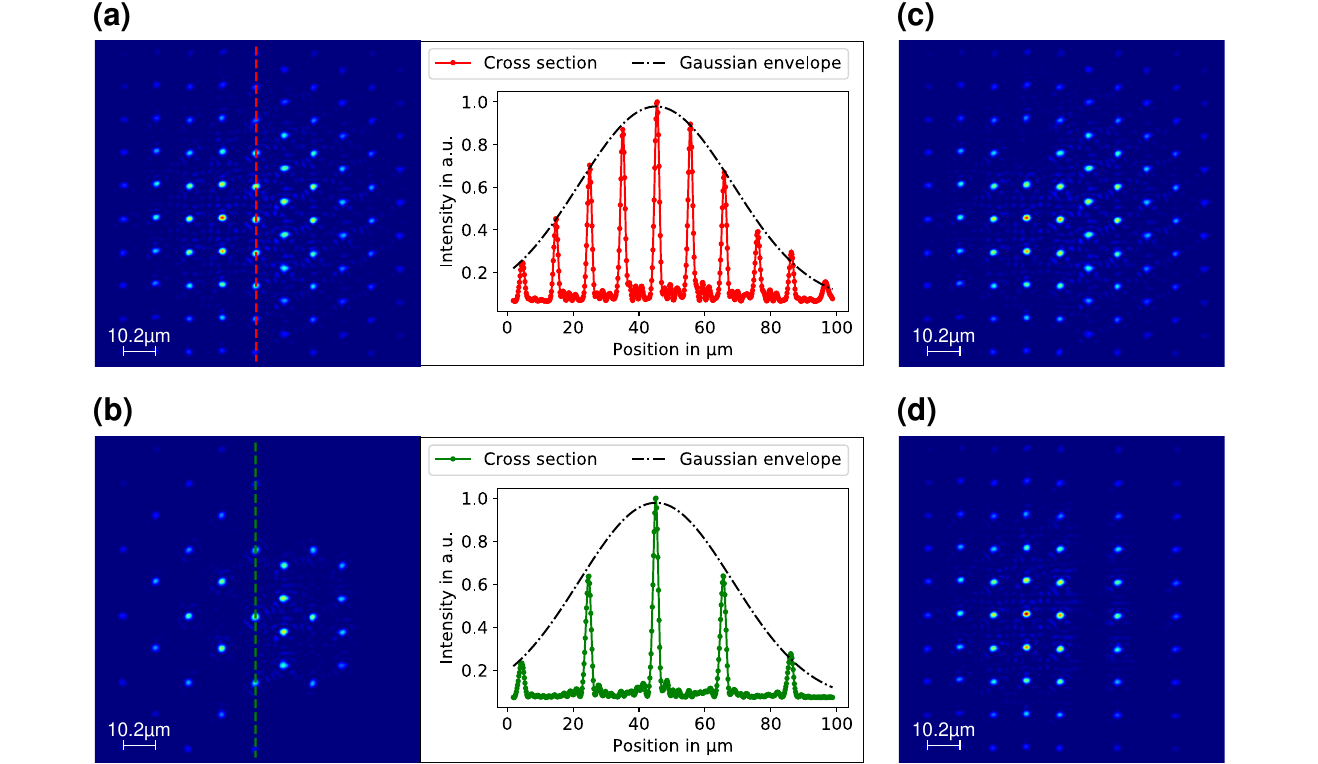}
	\caption{
		Characterization of the tweezer array after reimaging:
		(a) Fully enabled tweezer array with 97 lenslets (left) and a cross section along the dashed line showing the diffraction-limited tweezer spots (right).
		For our application we illuminate the DMD with a Gaussian laser beam, resulting in a Gaussian envelope of the tweezer intensities.
		(b,c,d) Selective DMD addressing of each lenslet enables the creation of versatile tweezer patterns. For a geometry featuring sets of deactivated spots (b, left) we observe a high extinction ratio between activated and deactivated spots (b, right). Due to DMD control, single-site defects (c) can be implemented as well as patterns with modified symmetries (d).
		\label{fig_geometries}}
\end{figure}
In this straightforward fashion, the setup grants access to versatile trap geometries of the tweezer array (Fig.~\ref{fig_geometries}) consisting of well defined foci with measured waists of $w_\mathrm{q}=\SI{1.36(5)}{\micro\meter}$ in the quadratic and $w_\mathrm{h}=\SI{1.29(6)}{\micro\meter}$ in the hexagonal region which are diffraction limited by the numerical aperture of the microscope objective L4 ($\mathrm{NA}=\SI{0.25(2)}{}$).
While in Fig.~\ref{fig_geometries}(a, left) an image of the fully activated array of 97 tweezers is shown, several tweezer spots are turned off in Fig.~\ref{fig_geometries}(b, left), in order to generate a modified periodic pattern.
The accessible tweezer geometries range from the full array to selected sections of pure quadratic or hexagonal symmetry, to the implementation of single-site defects through deactivation of single spots (Fig.~\ref{fig_geometries}(c)), and to freely definable symmetries (Fig.~\ref{fig_geometries}(d)).
The cross sections (Fig.~\ref{fig_geometries}(a, right) and Fig.~\ref{fig_geometries}(b, right)) confirm the high quality of the tweezer pattern and the high contrast at which individual tweezers can be switched.
The Gaussian envelope extracted from a fit to the cross section in Fig.~\ref{fig_geometries}(a, right) is consistent with the shape of the beam incidenting on the DMD. 
If a uniform intensity distribution for all tweezers is required \cite{Sturm2017}, either an incident beam with flat top profile can be used or site-selective intensity optimization via deactivating a subset of the mirrors corresponding to each lenslet can be applied. Due to the large number of \SI{730(8)}{} mirrors per lenslet, intensity adjustment can be achieved on a $\SI{0.2}{\percent}$ level in the range of \SI{20}{\percent} to \SI{100}{\percent} of the full intensity without degrading the quality of the tweezers.

Finally, we want to put these results into perspective with state-of-the-art tweezer arrays used for our targeted application. As demonstrated in \cite{OhldeMello2019}, microlens-array-based optical tweezer setups are routinely used to trap individual neutral atoms in foci with a waist of typically \SI{1.45(10)}{\micro\meter} and a power of about \SI{1}{\milli\watt} facilitating a trap depth on the order of $k_B\cdot\SI{1}{\milli\kelvin}$. The presented platform provides these parameters with additional means of flexibility. Furthermore, since the tweezer spot size is ultimately determined by the the NA of L4, the optical setup can be easily adapted to applications such as confocal or super-resolution microscopy at the limits of current technology by using a state-of-the-art high-NA microscope objective as L4.
\section{Conclusion}
\label{sec_conclusion}
We have introduced a novel universal platform for the creation of large-scale tweezer arrays with single-site addressability based on microlens arrays produced by 3D femtosecond direct laser writing. We have manufactured an array composed of 97 circular lenslets in quadratic and hexagonal packing and create a tweezer array with foci having diffraction-limited waists in the single-micrometer regime and separations of \SI{10.2(3)}{\micro\meter}. Through modification of our relay optics, the tweezer parameters of the reported platform can easily be adapted to other tweezer implementations with state-of-the-art parameters and added flexibility.
As each tweezer is defined by the corresponding microlens within a solid optical microlens array, the present system exhibits a high degree of stability and precision, which would typically be at the expense of short-term reconfigurability. In contrast, our system is perfectly suited for on-demand optimization of the imprinted optics via rapid prototyping and exhibits additional means of dynamical reconfiguration owing to the DMD-based spatial control of the incident light. The achieved efficiency for our DMD-controlled MLA of about \SI{25}{\percent} is below the values of \SI{80}{\percent} typical for holographic usage of phase modulating SLMs while exceeding the efficiency of holographic setups relying on DMDs which is typically at about \SI{10}{\percent}. The large number of mirrors assigned to each lenslet also renders gradual trap depth modulation possible via partial deactivation of mirrors. Moreover, the present system can be easily extended by an ancillary movable optical tweezer in the manner detailed in \cite{Barredo2018,OhldeMello2019} for the purpose of transporting objects between traps or equipped with parallelized position control as demonstrated in \cite{Schlosser2011}. Additionally, our approach is not constrained to two dimensional arrays: imprinting lenses with a range of different focal lengths makes versatile three-dimensional geometries accessible in a straightforward way.
\section*{Funding}
Bundesministerium f\"{u}r Bildung und Forschung (Printoptics); Baden-W\"{u}rttemberg Stiftung (Opterial); European Research Council (PoC 3D PrintedOptics); Technische Universit\"{a}t Darmstadt (OA Fund); Deutsche Forschungsgemeinschaft (SPP 1929 GiRyd BI 647/6-2).

\section*{Disclosures}
The authors declare that there are no conflicts of interest related to this article.
      
\bibliography{Printed_MLA} 

\end{document}